# A Mach-Zehnder interferometer based tuning fork microwave impedance microscope

Z. Liu, P.W. Krantz, and V. Chandrasekhar*

*Department of Physics, Northwestern University, Evanston, Illinois. 60208, USA*

We describe here the implementation of an interferometer-based microwave impedance microscope on a home-built tuning-fork based scanning probe microscope (SPM). Tuning-fork based SPMs, requiring only two electrical contacts for self-actuation and self-detection of the tuning fork oscillation, are especially well suited to operation in extreme environments such as low temperatures, high magnetic fields or restricted geometries where the optical components required for conventional detection of cantilever deflection would be difficult to introduce. Most existing and commercially available systems rely on optical detection of the deflection of specially designed microwave cantilevers, limiting their application. A tuning-fork based microwave impedance microscope with a resonant cavity near the tip was recently implemented: we report here an enhancement that incorporates a microwave interferometer, which affords better signal to noise as well as wider tunability in terms of microwave frequency.

## I. INTRODUCTION

The tip of a scanning probe microscope (SPM) can be modified in a number of ways to enable different microscopy modes to probe a range of properties of samples of interest.[1] For example, coating the tip with a magnetic film probes the local magnetic properties of a sample; applying a voltage to a conducting tip probes the electrical properties. Measuring a sample's response with electromagnetic radiation emanating from the microscope tip enables probing the properties of a sample on a scale determined by the dimensions of the tip, typically far less than the wavelength of the applied radiation. A well known example of such near field scanning microscopy is near-field scanning optical microscopy (NSOM),[2] where the near field scanning resolution is determined by the aperture of an optical fiber of the order of a few tens of nanometers, far smaller than the 100s of nanometers corresponding to the wavelength of the applied radiation.

Our interest is in microwave impedance microscopy (MIM),[3] in which a microwave signal is applied to the microscope tip. Here again the scanning resolution is determined by the size of the tip, in the range of 10-50 nm, and not by the millimeter scale wavelength of the applied radiation. The advantage of MIM is that it makes available a non-contact means of determining the electrical properties of structures with high resolution, allowing, for example, the visualization of buried conducting parts of devices. Scanning MIMs are now commercially available, and are used to detect defects in semiconductor devices not easily seen by other means. With appropriate modeling, quantitative estimates of the real and imaginary part of the conductivity can be obtained over a wide frequency range.[4–10] MIM is also a powerful probe for fundamental studies of new phenomena: MIM has been used to image quantum Hall edge stages[11, 12] and quantum spin Hall edge states in thin film devices[13], as well as an axion-insulator state at the edge of two films[14].

Microwave electronics operates best when all elements of a microwave circuit have the same impedance, typically 50 Ω. A significant problem in MIM is matching the impedance of the microscope components to the impedance of the microwave detection circuit. A major effort has been made to fabricate special cantilever tips with integrated coplanar waveguides right down to the conducting tip of the cantilever in order to optimize the coupling to the microwave detection circuit,[15, 16] but the unavoidable interface between the tip and the sample invariably represents an impedance that is 10s of MΩ, and thus circuits are used near the tip to match the tip-sample impedance to the characteristic impedance of the microwave detection circuit to improve the signal to noise ratio.[17–19]

The simplest such circuit is a $\lambda/2$ resonator made by attaching a few-cm length of coaxial cable to the tip, with the other end connected to the microwave detection circuit through a small (~0.1 pF) capacitor, with or without a 50 Ω terminator placed in parallel at the interconnect. The reflection signal ($S_{11}$) from this arrangement shows deep dips at frequencies corresponding to the fundametal and harmonic frequencies of the resonator. No reflection corresponds to a 50 Ω impedance as observed by the microwave detection circuit, so that the deeper the dips in $S_{11}$, the better the impedance matching. At these resonant frequencies, the detection circuit is most sensitive to changes in the tip-sample capacitance that lead to changes in the resonant frequency and the magnitude of $S_{11}$. Such matching circuits are now used in commercial instruments with the aforementioned specially fabricated microwave cantilever tips, but also work surprisingly well in tuning-fork SPMs where the tip is fabricated from a bare, unshielded wire a few millimeters in length connected to the central conductor of the coaxial cable of the resonator, and etched at the other end to form the tip.[20] Another scheme that was recently implemented[21] is a specially fabricated miniature superconducting resonator attached to the end of a tuning fork that is inductively coupled to a microwave strip line, allowing operation at very low microwave powers.

---

*Electronic address: v-chandrasekhar@northwestern.edu

One disadvantage of using a fixed $\lambda/2$ resonator as described above is that one is restricted to measurements at the fundamental and harmonic frequencies of the resonator. For a resonator of length ~10 cm the dips in $S_{11}$ occur at multiples of 1 GHz, so that one is unable to scan at intermediate frequencies without changing the length of the coaxial line forming the resonator. Recently, a number of groups[22–25] have demonstrated interferometer-based microwave detection circuits coupled to coplanar microwave cantilevers that does not use a special resonator placed close to the tip, but instead uses the entire length of microwave connection from the microwave detection electronics to the tip sample interface as one arm of an effective Mach-Zehnder interferometer (MZI), matched against another arm whose attenuation can be tuned. One again sees dips, now in the transmission ($S_{21}$) of the interferometer, but since the length of the relevant "resonator" is now the length of the entire line which can be a meter or two in length, the dips are much more closely spaced in frequency, and by implementing a phase shifter in the local arm of the interferometer, the position of the dips can be tuned essentially continuously in frequency. This detection circuit has been shown to have better a signal-to-noise ratio (SNR) than the $\lambda/2$ resonator described earlier with commercial MIM tips.[25]

We describe below the implementation of such a Mach-Zehnder interferometer based MIM on a compact homebuilt tuning-fork SPM using an etched tungsten wire as a tip. A key ingredient is being able to demodulate the direct MIM signal at the frequency of oscillation of the tuning fork using a lock-in amplifier. We demonstrate its perfomance by imaging conductivity variations in degraded thin metal films deposited on oxidized silicon wafers.

## II. DESCRIPTION OF THE MICROWAVE ELECTRONICS

Figure 1 shows a schematic of the microwave electronics. The heart of the instrument is the Mach-Zehnder interferometer, which is simply a commercial 90-degree hybrid coupler (Minicircuits ZX10Q-2-19-S+ [26]). The hybrid coupler splits the signal applied at its input into two signals phase-shifted from each other by 90 degrees.[27] One of these signals (the port labeled "0" in the figure) feeds the transmission line that is connected through coaxial cables to the tip attached to the tuning fork. The other signal (the port labeled "90" in the figure) is connected to one port of a voltage variable attenuator (VVA, Minicircuits EVA-3000+) whose attenuation can be controlled in the range 3-26 dB by means of a dc voltage applied to one of its terminals. Reflections from these two ports are combined in quadrature at the 4th port of the hybrid coupler (labeled "Iso" in the figure). By varying the attenuation of the VVA, the signals from the two paths can be made to interfere destructively, leading to deep dips in the signal out from the Iso port of the hybrid coupler at specific frequencies determined primarily by the path length between the "0" port and the tip mounted on the tuning fork. This path length is typically 2 m or longer, giving dips at the Iso port at intervals of approximately 50 MHz. As mentioned earlier, the resonance frequencies at which the dips occur can be tuned essentially continuously by placing a phase shifter in one arm of the Mach-Zehnder interferometer, although we have not implemented this feature in the current instrument. Note, however, that the frequency range of operation is limited by the bandwidth of the hybrid coupler. The one used here (Minicircuits ZX10Q-2-19-S+) has an operating frequency range of 1.1-1.925 GHz, but the frequency range can be changed readily by swapping in a different hybrid coupler (the ones from Minicircuits have the same form factor) or using the more expensive option of utilizing a coupler with a wider bandwidth.

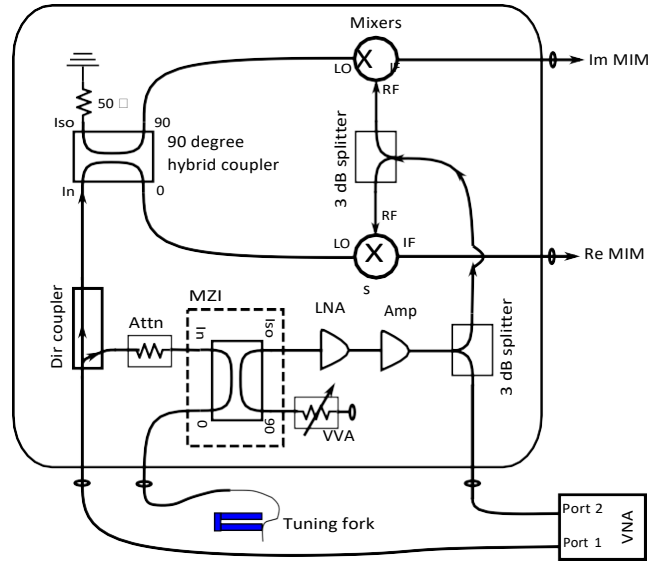

FIG. 1: Overall schematic of the microwave detection circuit.

The input port of the MZI is driven by a microwave signal sourced from Port 1 of a vector network analyzer (VNA, Agilent 8753 [28]) through a directional coupler. The through port of the directional coupler feeds a second 90-degree hybrid coupler that is used in a conventional manner to provide two signals in quadrature to drive the local oscillator (LO) ports of two mixers (Minicircuits ZX05-C24-S+). These are so-called Level 7 mixers, i.e., they require 7 dBm of power at their LO ports for optimal operation. Since the directional coupler and 90-degree hybrid coupler each introduce a 3 dB loss, the VNA provides 13 dBm of power. Consequently, an attenuator may be placed between the directional coupler and the input port of the MZI in order to limit the power ap-plied. The output of the MZI is amplified through a series of amplifiers, the first stage of which is a low noise amplifier (LNA). The amplified signal is then sent through a 3 dB splitter, one output of which is connected to Port 2 of the VNA for $S_{21}$ measurements. The other output of the splitter is fed to the radiofrequency (RF) ports of the mixers through a second 3 dB splitter so that one can simulaneously monitor the response at the VNA as well as the mixers. For optimal operation of the mixers the signal applied at the RF ports is recommended to be no more than 10 dB down from the LO signal for optimal perfomance, so the gain of the amplifier(s) must be sufficient to provide this. Finally the intermediate frequency (IF) outputs of the mixers are provided as analog outputs, where they can be read directly by the SPM electronics after appropriate filtering.

In principle, one does not require both a VNA and the mixers. One could read the $S_{21}$ signal directly from the VNA, or just use the mixers with an independent microwave source. Unfortunately, the VNA at our disposal does not provide an output corresponding to $S_{21}$ that can interface with our SPM electronics with a high enough update rate for scanning. In addition, we obtain greater sensitivity and immunity to drift and noise if we demodulate the outputs with a lock-in amplifier at the frequency of oscillation of the tuning fork, typically 30 kHz. Consequently, it is very useful to have analog outputs with sufficiently fast response, which is provided by the IF outputs of the mixers. On the other hand, the VNA proves to be very useful in quickly identifying and tuning the resonances with the VVA, and in ensuring that the microwave amplifiers are not damaged during this process, as we discuss below.

wire tip by itself, as it would be if we were ready to scan (the tip is far from the sample). Oscillations are clearly seen with a period of about 50 MHz, corresponding to resonances of the entire length of the assembly from the VNA to the tip of the etched W wire attached to the tuning fork. However, the amplitude of the oscillations is only about 1 dB on a logarithmic scale. For comparison, Fig. 3(a) shows the transmission spectrum ($S_{21}$) of the MZI with the same coaxial line as one arm of the MZI measured with the VNA alone (no amplifiers), with no attenuator in the circuit, and with the VVA adjusted to obtain the largest amplitude of oscillation. (The frequency range for both plots is restricted to be between 1-2 GHz, corresponding to the bandwidth of the 90-degree hybrid coupler used in the MZI.) Figure 3(a) shows that the resonances are now much deeper, about 45 dB in amplitude. In fact, this is due to the finite frequency resolution of the VNA over the scan range: if we zoom in on a specific resonance as shown in Fig. 3(b), we see that the oscillation amplitude can be larger than 80 dB relative to the maxima in Fig 3(a). This should be compared to the approximately ~30 dB oscillation amplitude reported for the $\lambda/2$ resonators discussed above.[25]

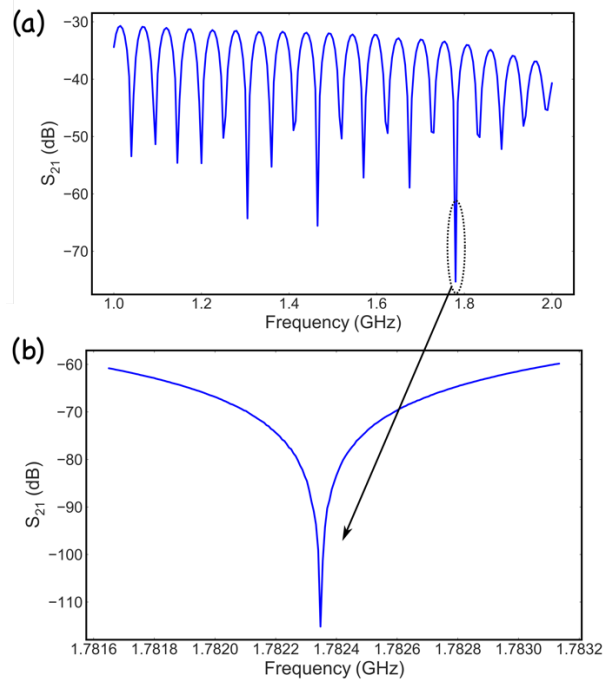

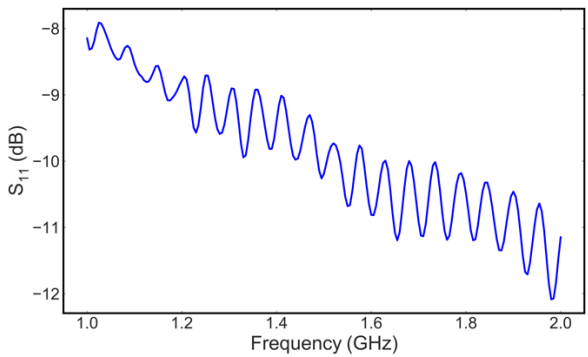

FIG. 2: Reflection spectrum of the coaxial line connected to the SPM tip by itself.

FIG. 3: (a) Transmission $S_{21}$ spectrum of the MZI over the frequency range 1-2 GHz. (b) Zoomed in spectrum of one resonance, showing the true depth of the resonance dip.

### A. Resonance characteristics

Figure 2 shows the reflection spectrum ($S_{11}$) of the coaxial line connected through the microscope to the W

### B. Choice of amplifiers

The $S_{21}$ spectrum shown in Fig, 3(b) is relative to the input power provided by the VNA. If we source the VNA

at 13 dBm in order to optimize the performance of the mixers as discussed above, the amplitude of the signal at a dip similar to the one shown in Fig. 3(b) would be ~-100 dBm. Consequently, the microwave amplifier(s) in the circuit should ideally provide a gain of 90-100 dB to supply the mixers with a sufficiently large signal. To provide such a large gain, we use a number of amplifiers in series. As with all such arrangements, the first amplifier should have the lowest noise with reasonable gain. We use a Minicircuits ZX60-P33ULN+, which has a gain of approximately 15 dB and a noise figure of 0.38 dB in the 1-2 GHz range of operation here. The second stage amplifier is a Minicircuits ZX60-2534MA-S+ with 43 dB of gain. For the third stage, we have used either a Minicircuits ZX60-V63+ with 21 dB of gain, for a total of 79 dB gain, or a MITEQ AM-4A-1020 with 40 dB of gain,[29] for a total of 98 dB of gain. The choice of amplifiers clearly depends on the frequency bandwidth of interest: the most critical amplifier is the first stage amplifier, which should have the lowest noise.

Given the sharpness of the resonances and the high gain of the amplifier stage, care must be taken in tuning the system to a particular resonance dip in order not to damage the amplifiers. For example, the maximum power that can be applied to the input of the second stage amplifier mentioned above (Minicircuits ZX60-2534MA-S+) is -15 dBm per its specifications. In reality, the 1 dB compression point for this amplifier is 18 dBm at the output, so -25 dBm is the maximum power that can be applied to its input without distorting the signal given its gain of 43 dB. It would seem that this is not a problem considering the -100 dBm levels of the signals at the resonance dip mentioned above. However, the sharpness of the resonances means that the signal levels rise rapidly around the resonance frequency. If in searching for a suitable resonance, the frequency range is not narrow enough, it is easy to cause distortion or even inadvertently damage one or more later stage microwave amplifiers, as we have done at least once. It is here that the VNA proves very useful: We start by sourcing a very low power (-60 dBm) from the VNA in order to choose a suitable resonance over the entire operating frequency range (here 1-2 GHz). We then progressively narrow the frequency range around the chosen resonance while increasing the VNA source power and tuning the resonance with the VVA until the source power is at the desired level of 13 dBm. At this point, we set the VNA on continuous wave (CW) mode at the resonance frequency, and can use the IF outputs of the mixers as inputs to the scanning probe microscope directly, or through a lock-in amplifier for demodulation at the tuning fork resonance frequency.

### C. Mixers

After amplification and losses due to the 3 dB splitters, the magnitude of the signals presented at the RF ports of both mixers (Minicircuits ZX05-C24-S+) is in the range of -45 to -25 dBm with the two amplification stages discussed above. The conversion loss of these mixers with a 7 dBm input at the LO ports is about 6 dB, so that the output signals at the IF ports of the mixers are in the range of roughly -50 to -30 dBm, or 1-10 mV rms. As one normally tunes the microwave electronics with the tip far away from the sample, this is the magnitude of the signal in this situation. As the tip moves closer to the sample, the resonance shifts, and consequently the amplitude of the signal seen at the IF ports increases (see below), but its average value is still in the range of a few tens of millivolts.

Nominally, the 90-degree hybrid splitter feeding the LO ports of the two mixers in quadrature (see Fig. 1) enables one mixer to measure the component of the interferometer signal that is in-phase (i.e., the "real" component) and the other mixer to measure the out-of-phase (imaginary) component. In reality, however, there is an arbitrary phase shift introduced by the path length of the interconnecting cables that changes with frequency. One could place a phase shifter in the path feeding the 90-degree hybrid and adjust the phase to minimize/maximize the outputs of the mixers as is done in any phase sensitive measurement. However, such phase shifters are expensive and have typically have very limited bandwidth. We instead choose the channel to measure by monitoring both signals on an oscilloscope as we perform coarse approach, and selecting the channel that shows the largest change on approach. This is a subjective selection as sometimes both channels show changes of the same order. For more quantitative analyses of the resistive and reactive response of a sample, it is of course important to be able to tune the phase of the mixers so that one channel represents the in-phase and the other channel the out-of-phase component of the response.[3]

### D. Demodulation at the tuning fork oscillation frequency

The signals at the IF outputs of the mixers can be read directly by our home-built SPM control program and electronics after appropriate low pass filtering. However, as noted by other groups,[16, 20] these signals are subject to drift, and in our case the changes in them are quite small. In order to pull the relevant signal out, we demodulate either mixer output at the frequency of oscillation of the tuning fork. We use a venerable EG&G PAR 124 lock-in amplifier referenced to the drive of the tuning fork, which in our case is provided by a home-built tuning fork controller. The PAR 124 is preferred over the digital lock-in at our disposal (an EG&G 7260 [30]) because it has a narrow input bandpass filter that can be precisely tuned to the reference frequency and it provides an analog voltage output with a fast enough time constant (<1 msec) that can then be read by our SPM control program. The lock-in is phased to maxi-mize the signal to be read once the tip has approached prior to scanning, as the signal is essentially zero when the tip is more than a few microns away from the surface. Sometimes the signal is still very small even after close approach if one lands on a non-conducting region of the sample, in which case the phase will need to be adjusted after identifying and moving to the conducting regions of the sample.

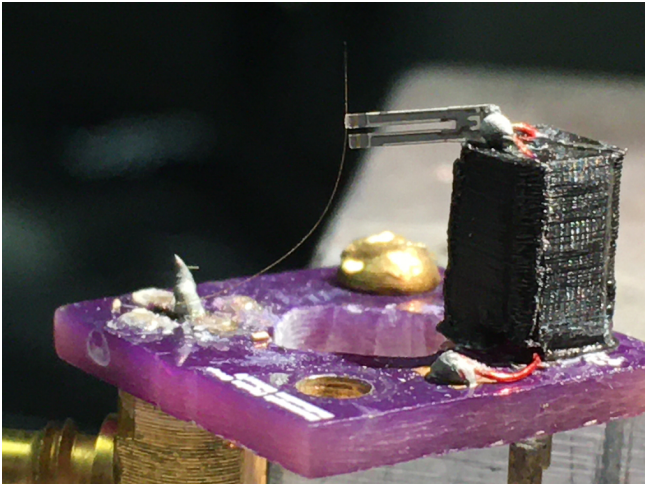

FIG. 4: Image of the tip and tuning fork stage. The square PCB stage is 13 mm on a side.

### E. Tuning fork stage

Figure 4 shows an image of the tuning fork and tip stage. The tip and tuning fork are mounted on a custom printed circuit board (PCB) on which a microwave SSMC PCB launch (partially visible on the lower left of the image) and two pins for connection to the tuning fork (partially visible on the lower right) are soldered. Mounting tuning forks and etching tips have been described by many groups including ours,[31] and will not be discussed here. We only note here the special considerations for MIM. The 25 $\mu$m Au coated W wire is soldered to the central pin of the SSMC PCB mount. In our other (non-MIM) SPM stages, the tuning fork is mounted directly on the PCB. However, the diameter of the central pin on the SSMC connector is a few millimeters and might dominate the capacitance between the central conductor and the sample. Hence, we place the tuning fork on a 6 mm long pillar so that the wire tip is almost 10 mm closer to the sample than the central pin of the SSMC connector.

Since the entire length of the line from the microwave interferometer electronics to the tip of the wire forms the resonator, it is critical that losses in the lines be minimized. The last section of a few millimeters of bare W wire and the reflections from the soldered connection to the SSMC launch cannot be avoided, but it is essential that lines and connections up to the PCB stage be of high quality. Consequently, we use semi-rigid coax throughout: UT-85 cable from the microwave electronics to the SPM box, and smaller diameter semi-rigid coax for the connections inside the SPM box.

The tuning-fork/tip PCB stage is mounted along with the sample to be imaged on our home-built SPM.[32] For an appropriate transducer signal for feedback control, one can use the amplitude of oscillation, phase of oscillation or frequency of oscillation of the tuning fork. The quality factor $Q$ of the tuning fork with the tip mounted ranges from 500-1000 at room temperature in air. For these studies, we have therefore used amplitude feedback. In other regimes such as in vacuum and at low temperatures where the $Q$ may rise to 30,000-100,000, one can use frequency feedback with a phase-locked loop (PLL) or $Q$ control to reduce the $Q$. All these modes can be implemented by our home-built tuning fork controller.[33]

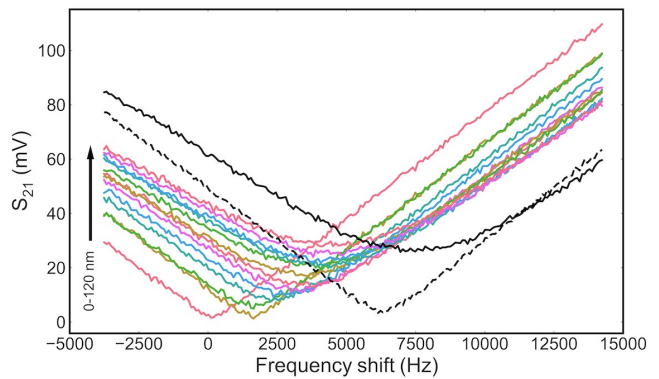

FIG. 5: $S_{21}$ measurement of a resonance dip on a linear scale as a function of distance. The colored traces correspond to progressive increases of the tip height by 10 nm, from 0 nm (after immediately stopping feedback) to 120 nm using the Z piezo. The solid black trace is when the tip is retracted by approximately 6 $\mu$m using the coarse approach stage. The dashed black line shows the initial tuning of the spectrum with the VVA in this retracted position, demonstrating the drift of the direct signal over a span of about 30 minutes. The $x$-axis corresponds to frequency shifts from a center frequency of about 1.369 GHz.

## III. EXPERIMENTAL RESULTS

Figure 5 shows the $S_{21}$ spectrum of the MZI on a linear scale around a resonance dip of about 1.369 GHz as a function of the distance of tip from the surface, with no attenuator between the source and the MZI and a total gain of 98 dB in the amplifiers. To obtain the curves in Fig 5, we first performed a coarse approach to bring the tip in close proximity to the sample in feedback, turned off the feedback, and retracted the tip from the surface by about 6 $\mu$m using the Z coarse approach. The resonance was then tuned using the VVA in this position to obtain the dashed black curve in the figure. Experimentally, it turns out that the signature of optimal tuning of the MZI is the sharp 'V' shaped spectrum seen in this curve. The tip was then brought to the surface under feedback, resulting in a shift in the position of the resonance dip down by about 6 kHz. The feedback was then turned off, and the spectrum taken as the tip was retracted in 10 nm steps from 0 to 120 nm, resulting in the colored traces shown in the figure. The tip was then retracted to the nominal starting point (given the hysteresis in the coarse approach) and the final solid black spectrum was recorded.

There are a number of things to note about this plot that inform the protocol we use to scan. First, the amplitude of the signal is small, on the order of a few millivolts at the minima, even with 10 dBm of power applied to the input port of the MZI. This is the signal level expected given the sharpness of the resonance dips (see Fig. 3(b)) and the gain of the amplifiers. Other groups have reported output signals on the order of volts,[20] but this is with the $\lambda/2$ resonator where the amplitude of the resonance is 30-40 dB, and the total amplifier gain of the order of 110 dB. We could increase the signal level to the same range with an additional 30-40 dB of gain, but the gain of 98 dB used for obtaining the plots in Fig. 5 is sufficient for obtaining good images. Second, the shift in frequency from the retracted position to where the system is in feedback with the tip at the sample is small. Since tests of the instrumentation with open end of a semi-rigid coax approaching a metallic surface show much larger changes, we attribute this to the termination of the coaxial line with the bare W wire with a small tip size (typically 50 nm). (Other groups see large frequency shifts with shielded probes.[18]) As noted earlier, we adjust the VVA to obtain the deepest minimum and fix the frequency at this minimum when the tip is retracted. As the tip approaches the sample, the resonance minimum will shift, and the magnitude of the signal at the fixed frequency will rise to a few 10s of millivolts. Finally, we can see that there is a drift in the spectrum as a function of time. This is most evident in the shape of the curve, which goes from the sharp 'V' shape to something more parabolic at the minima, in addition to the expected shift in frequency with distance. The difference is clearly seen between the initial retracted curve (dashed black line) and final retracted curve (solid black line), with the solid black curve being taken approximately 30 minutes after the dashed black curve. The drift in the direct signal has been reported earlier,[16, 20] and is attributed to temperature changes in the microwave electronics, slight changes in connectors, etc. In our case, we also find that the position of the minimum of a particular resonance is extremely sensitive to the adjustment of the VVA, hence any changes in the voltage applied to the VVA through changes in temperature or drift of the biasing circuit will also result in drift of the output signal. Consequently, using the filtered IF outputs of the mixers directly as inputs to the SPM program will result in images with drift.

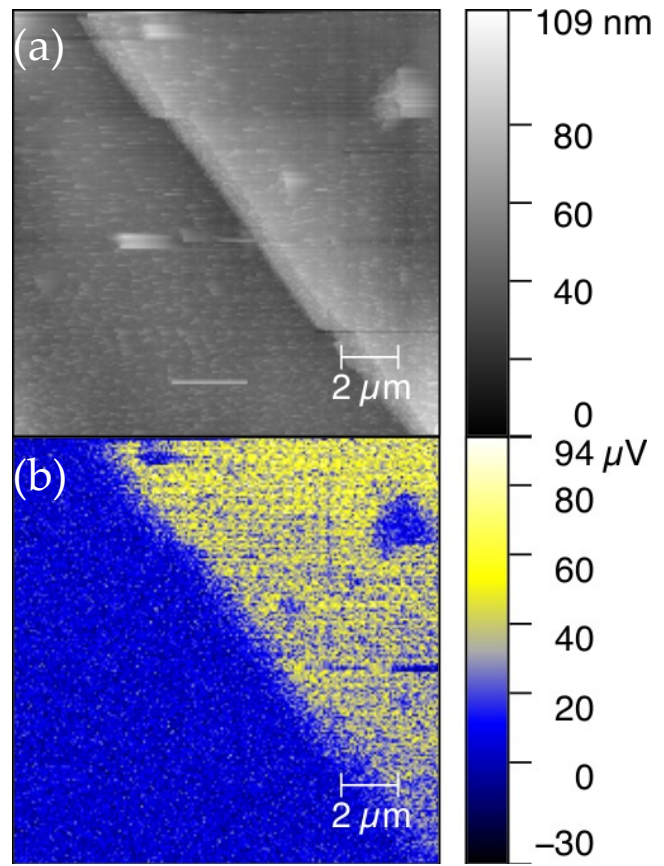

FIG. 6: Large area topographic scan (a) and simultaneous MIM image (b). The MIM image corresponds to the demodulated output of the IF signal of the mixer at the frequency of oscillation of the tuning fork. The frequency of the resonance is 1.7825 GHz and the total amplification of the signal from the MZI is 79 dB.

In order to circumvent this problem, we take advantage of the fact that the tuning fork is oscillating at a frequency of 30 kHz, as has been done by Cui *et al.*[20] During each cycle of oscillation, the tip moves closer and farther away from the sample, changing the signal at the same frequency. Demodulating the IF outputs of the mixers at the frequency of oscillation of the tuning fork results in a signal that is sensitive to only the changes associated with the oscillation, and hence is much less sensitive to the overall drift in the IF outputs. In order to demodulate the signal, we reference the PAR 124 lock-in amplifier to the oscillator driving the tuning fork, tune the bandpass of the input filter of the lock-in to the reference signal, then phase the signal to obtain the full in-phase response.

To demonstrate the performance of the MIM, we imaged a sample fabricated in our group. This sample consists of Au contact pads and large area leads fabricated by photolithography, and finer features of Au and Al fabricated by electron-beam lithography on an insulating Si substrate with 1 μm of silicon oxide. The unusual characteristic of this sample is that it had been sitting on a shelf under ambient conditions for more than a decade, so that the metallic films deposited on the substrate have deteriorated over time, and could be anticipated to show variations in conductivity.

Figure 6(a) shows a 15 μm x 15 μm topographic scan of the edge of an Au contact lead on the sample, and Fig. 6(b) shows the corresponding MIM image taken simultaneously. The MIM image is from the demodulated output of the lock-in amplifier as described above. The metal film is at the top right of the image, and it can be seen that in this region, the amplitude of the MIM signal is larger, while on the oxide surface at the lower left, the MIM signal essentially vanishes. The sample has debris on its surface that appear to be insulating from the fact that the MIM signal vanishes when the tip is on the debris. The magnitude of the MIM signal is small, on the order of 100 μV, a result of the relatively small total amplification (79 dB) as well as the small amplitude of oscillation of ~0.2 nm of the tuning fork (based on interferometry measurements that we have performed on similar tuning forks).

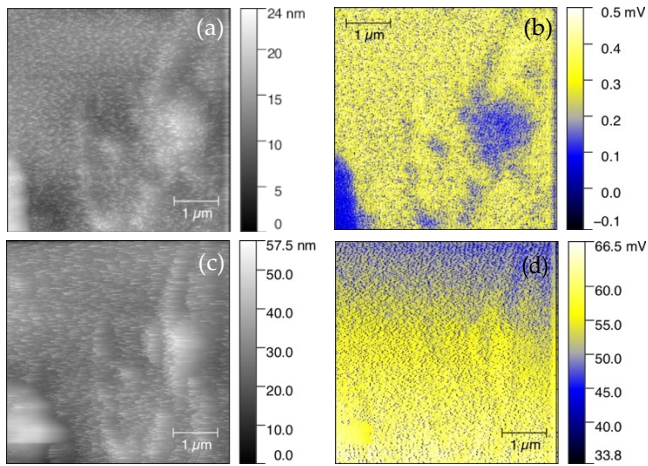

FIG. 7: Comparison of direct mixer output and demodulated mixer output. (a) Topographic scan and (b) simultaneous lock-in MIM image. (c) Topographic scan and (d) simultaneous direct MIM image. Features in (d) are barely discernible, and the drift in the signal from the top of the image to the bottom is clearly evident. The frequency of the resonance is ~1.359 GHz and the total amplification of the signal from the MZI is 98 dB.

In order to demonstrate the improvement in contrast and drift of the demodulated signal in comparison to the the direct IF output of the mixers, Fig. 7 shows successive 5 μm x 5 μm simultaneous topographic/MIM scans of the same area of a metal contact pad, the first taken using the demodulated lock-in output and the second using the direct IF output of a mixer. In comparison to the demodulated signal, features in the direct signal are difficult to make out, and the general drift in the amplitude of the signal from the top of the scan to the bottom

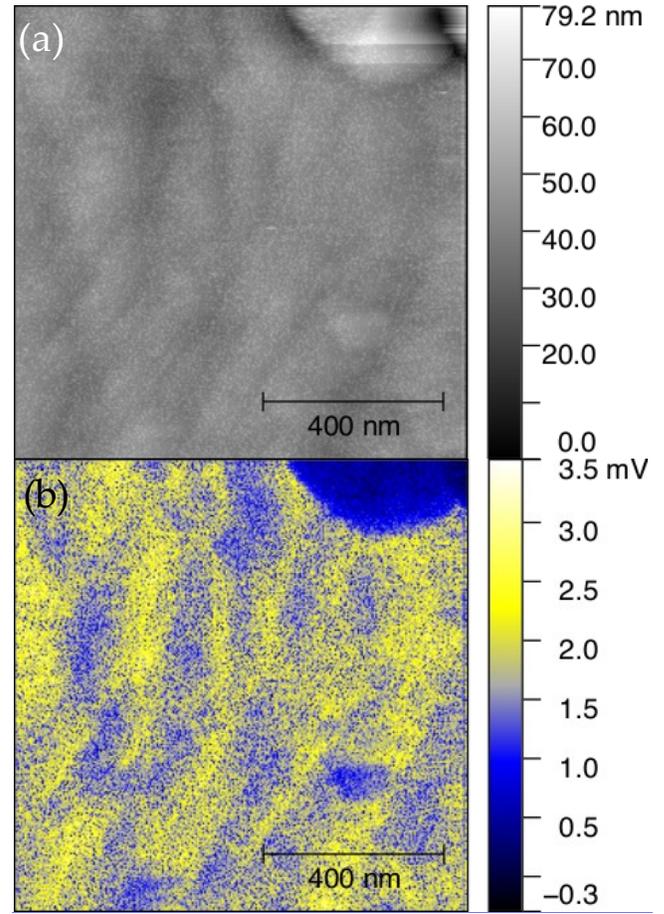

FIG. 8: Simultaneous small area topographic (a) and MIM (b) image of a metallic film The frequency of the resonance is 1.359 GHz and the total amplification of the signal from the MZI is 98 dB.

is clearly evident, a direct result of the drift in the $S_{21}$ signal seen in Fig. 5.

As a final example of the stability of the demodulated output, Fig. 8 shows the resulting MIM image of an approximately 6 hour scan of a 1 μ x 1 μm area of a contact pad. Apart from the piece of debris at the right top corner, the surface shows height variations of the order of 5 nm that we believe are due to the grain structure of the film. The grain structure is also reflected in the MIM image. The MIM image demonstrates that the instrument is able to distinguish small variations in conductivity associated with the aging of the metal film.

In summary, we have demonstrated an interferometer microwave impedance microscope based on a tuning fork SPM that is capable of imaging small variations in conductivity and can operate over a wide frequency range. The microscope is compact enough to be used in extreme environments such as low temperatures and high magnetic fields.

The data that support the findings of this study are available from the corresponding author upon reasonable request.

**Acknowledgments**

We thank Kevin Ryan for designing some parts of the PCB stage, and José Aumentado for discussions regarding the microwave electronics. This research was conducted with support from the US Department of Energy, Basic Energy Sciences, under grant number DE-FG02-06ER46346.

[1] S.V. Kalinin and A. Gruverman, Eds., *Scanning Probe Microscopy*, Springer (2007).
[2] See, for example, R.J. Hermann and M,J. Gordon, Annu. Rev. Chem. Biomol. Eng. **9**, 365 (2018).
[3] For a recent review, see Z. Chu, L. Zheng, and K. Lai, Annu. Rev. Mater. Res. **50**, 1.1 (2020).
[4] H. P. Huber *et al.*, Rev. Sci. Instr. **81**, 113701 (2010).
[5] H. Tanbakuchi, M. Richter, F. Kienberger, and H.-P. Huber, 'Nanoscale materials and device characterization via a scanning microwave microscope,' in 2009 IEEE International Conference on Microwaves, Communications, Antennas and Electronics Systems (IEEE, 2009), pp. 1-4.
[6] 2. O. Amster, Y. Yang, B. Drevniok, S. Friedman, F. Stanke, and St. J. Dixon-Warren, 'Practical quantitative scanning microwave impedance microscopy of semiconductor devices,' in 2017 IEEE 24th International Symposium on the Physical and Failure Analysis of Integrated Circuits (IPFA) (IEEE, 2017), pp. 1-4.
[7] T. S. Jones, C. R. Pérez and J. J. Santiago-Avilès, AIP Advances **7**, 025207 (2017).
[8] C. Gao, T. Wei, F. Duewer, Y. Lu and X.-D. Xiang, Appl. Phys. Lett. **71**, 1872 (1997)
[9] Chen Gao, Bo Hu, I Takeuchi, Kao-Shuo Chang, Xiao-Dong Xiang and Gang Wang, Meas. Sci. Technol. **16** 248 (2005).
[10] G. Gramse, M. Kasper, L. Fumagalli, G. Gomila, P. Hinterdorfer and F. Kienberger, Nanotechnology **25**, 145703 (2014).
[11] K. Lai *et al.*, Phys. Rev. Lett. **107**, 76809 (2011).
[12] Y. T. Cui *et al.*, Phys. Rev. Lett. **117**, 186601 (2016).
[13] Y. Shi *et al.*, Sci. Adv. **5**:eaat8799 (2019).
[14] M. Allen *et al.*, PNAS **116**, 14511 (2019).
[15] See, for example, K. Lai, W. Kundhikanjana, M. A. Kelly and Z.-X. Shen, Appl Nanosci **1**, 13 (2011).
[16] K. Lai, W. Kundhikanjana, H. Peng, Y. Cui, M. A. Kelly, and Z. X. Shen, Rev. Sci. Instr. **80**, 043707 (2009).
[17] S.M. Anlage, V.V. Talanov and A.R. Schwarz, in Ref.[1], pp.. 215-253 (2007).
[18] J. Kim *et al* Meas. Sci. Technol. **14**, 7 (2003).
[19] M. Tabib-Azhar, D. Akinwande, G. Ponchak and S.R. LeClair, Rev. Sci, Instr. **70**, 3381 (1999).
[20] Y.-T. Cui, E.Y. Ma and Z.-X Shen, Rev. Sci. Instr. **87**, 063711 (2016).
[21] S. Geany *et al.*, Scientific Reports **9**, 12539 (2019).
[22] T. Dargent, K. Haddadi, T. Lasri, N. Clément, D Ducatteau, B. Legrand, H. Tanbakuchi and D. Theron, Rev. Sci. Instr. **84**, 123705 (2013).
[23] K. Haddadi, S. Gu and T. Lasri, Sensors and Actuators A **230**, 170 (2015).
[24] G. Vlachogiannakis, H. T. Shivamurthy, M. A. D. Pino, and M. Spirito, " An I/Q-mixer-steering interferometric technique for high-sensitivity measurement of extreme impedances," in 2015 IEEE MTT-S International Microwave Symposium (IEEE, 2015), pp. 1-4.
[25] S.-S. Tuca, M. Kasper, F. Kienberger and G. Gramse, IEEE Trans. Nanotech. **16**, 991 (2017).
[26] Minicircuits, www.minicircuits.com.
[27] See D.A. Pozar, *Microwave Engineering*, 4th edition, Wiley [2012].
[28] Now Keysight Inc, http:\\www.keysight.com.
[29] This amplifier is no longer available.
[30] Now Ametek, www.ameteksi.com.
[31] P.W. Krantz and V. Chandrasekhar, J. Vac. Sci. Technol. B **38**, 024004 (2020).
[32] V. Chandrasekhar and M.M. Mehta, Rev. Sci. Instrum. **84**, 013705 (2013).
[33] V. Chandrasekhar, Rev. Sci. Instr. **91**, 023705 (2020).